\documentclass[twocolumn,footinbib,aps,prb]{revtex4-1}
\usepackage{color}
\usepackage{amsfonts,amssymb,amsmath,mathrsfs}
\usepackage{amsbsy}
\usepackage{graphicx}
\usepackage{hyperref}

\begin{document}

\title{Indications of a soft cutoff frequency in the charge noise of a Si/SiGe quantum dot spin qubit}

\author{Utkan G\"ung\"ord\"u}
\email{utkan@umbc.edu}
\affiliation{Department of Physics, University of Maryland Baltimore County, Baltimore, MD 21250, USA}

\author{J.~P.~Kestner}
\affiliation{Department of Physics, University of Maryland Baltimore County, Baltimore, MD 21250, USA}

\begin{abstract}
Characterizing charge noise is of prime importance to the semiconductor spin qubit community.  We analyze the echo amplitude data from a recent experiment [Yoneda et al., Nat. Nanotechnol. 13, 102 (2018)] and note that the data shows small but consistent deviations from a $1/f^\alpha$ noise power spectrum at the higher frequencies in the measured range. We report the results of using a physical noise model based on two-level fluctuators to fit the data and find that it can mostly explain the deviations. While our results are suggestive rather than conclusive, they provide what may be an early indication of a high-frequency cutoff in the charge noise.  The location of this cutoff, where the power spectral density of the noise gradually rolls off from $1/f$ to $1/f^2$, crucial knowledge for designing precise qubit control pulses, is given by our fit of the data to be around 200 kHz.
\end{abstract}

\maketitle

\emph{Introduction.}
A detailed understanding of the charge noise present in semiconductor spin qubit devices is vital for further progress towards using them as the basis for scalable fault-tolerant quantum computing. Spin qubits in silicon quantum dots \cite*{Burkard1999a,Culcer2009,Zwanenburg2013}
are a promising platform for quantum computation due to their electrical controllability and established fabrication techniques and existing facilities. The concentration of spinful isotopes in natural silicon can be reduced down to 800ppm or below with isotopic purification \cite*{Veldhorst2014a,Veldhorst2014c,Yoneda2017a,Takeda2018}, thus eliminating nuclear spin noise as a significant source of dephasing.
One-qubit gate fidelities close to \cite*{Veldhorst2014c,Watson2017a}  and exceeding \cite*{Yoneda2017a} 99.9\%  have been reported in both SiMOS and Si/SiGe devices.
Two-qubit gates, which rely on the electrically controlled exchange interaction, have also been experimentally demonstrated in SiMOS \cite*{Veldhorst2014a,Huang2018b} and Si/SiGe \cite*{Watson2017a,Zajac2017a,Russ2017a,Xue2018} spin qubits, but the gate fidelities of entangling operations are under 95\%, limited presumably by the charge noise which causes fluctuations in the exchange energy \cite*{VanDijk2018}.
Thus the presence of charge noise remains as one of the most important obstacles to realization of a fault-tolerant silicon based quantum computer with error rates below the $\sim$1\% threshold required for surface codes, let alone more demanding types of logical codes.

Despite being present in a wide range of condensed matter systems and its important role in limiting qubit coherence times, the origin of charge noise is still not completely understood \cite*{Paladino2014}. Measurement of the noise power spectral density (PSD) is an essential step toward understanding the nature of the charge noise, as well as designing higher-fidelity quantum gates. Even though dynamical decoupling (DD) is a tool used mainly for suppressing quasistatic noise \cite*{Vandersypen2005,Tomita2010a,Kabytayev2014}, it is also useful for probing noise spectra \cite*{Vandersypen2005,Biercuk2011,Alvarez2011,Bylander2011,Almog2016,Norris2016a,Sinitsyn2016,Paz-Silva2016,Szankowski2017}.

Although measurements of charge noise at low frequencies suggest SiMOS compares favorably
to Si/SiGe \cite*{Takeda2013,Freeman2016a}, measurements at higher frequencies in SiMOS devices have so far only been performed in the 1kHz-50kHz range \cite*{Chan2018a}. In Si/SiGe devices, on the other hand, the PSD has been probed up to 1MHz in Ref.~\onlinecite*{Kawakami2016}, but the data is inconclusive as to whether the tail of the PSD behaves like $1/f$ or $1/f^2$.
A recent experiment in Si/SiGe shows that the noise spectrum obeys the power law $\propto 1/f^{1.01}$ \cite*{Yoneda2017a} up to 320kHz.
In this paper, we analyze the data from this experiment and note that, despite the remarkably good agreement with a $1/f$ fit over a wide range of frequencies, there are systematic deviations from the power law noise model at high frequencies.  We show that these deviations might be explained by a standard noise model that assumes an ensemble of physical two-level fluctuators with a finite range of switching rates, in which case the PSD behaves as $1/f$ at intermediate frequencies and rolls off to $1/f^2$ at high frequencies.  By fitting the experimental data, we extract an estimated rolloff frequency.

\emph{Noise and qubit coherence.}
Noise with a $1/f^\alpha$ PSD
is a phenomenon which affects a wide range of systems, including the charge noise affecting spin qubits in silicon quantum dots. Although the exact mechanism responsible for this type of noise is currently unknown, it is commonly attributed to an ensemble of two-level fluctuators (TLFs) \cite*{VanDerZiel1950,Schriefl2006,Chirolli2008,Paladino2014,Muller2017}. Each such TLF switches between two states randomly at a rate $\gamma_i$, and the ensemble contributes to the Hamiltonian as
\begin{align}
H_\beta = \frac{1}{2}\beta(t)\sigma_z, \qquad \beta(t) = \frac{1}{2} \sum_i \hbar v_i \xi_i(t)
\end{align}
where $\xi_i(t) = \pm 1$ is due to stochastic fluctuations and
$\hbar v_i$ is the energy of the fluctuator.
Here, we assumed that TLFs mainly affect the qubit in the form of pure dephasing noise and that relaxation can be neglected, which is based on the fact that the $T_1$ of silicon spin qubits is much greater than their $T_2$ times.

The autocorrelation function of a single TLF is
\begin{align}
S_i(t) = \langle \hbar v_i \xi(t) \hbar v_i \xi(0) \rangle = (\hbar v_i)^2 e^{-|t| \gamma_i}
\end{align}
where $\gamma_i/2$ is the inverse switching rate of the fluctuator. This corresponds to the Lorentzian power spectrum
\begin{align}
S_i(\omega) = \int_{-\infty}^{\infty} dt e^{i\omega t} S_i(t) = 2\frac{ (\hbar v_i)^2  \gamma_i }{\omega^2 + \gamma_i^2}.
\end{align}

When the distribution function $P(v,\gamma)$ of a TLF ensemble depends exponentially on a physical quantity (such as a spatial distribution $\gamma \propto e^{l/l_0}$ with characteristic length $l_0$ or thermal distribution $\gamma \propto e^{-E/k_B T}$ where $E$ is the activation energy of a TLF), $P$ becomes log-uniform, i.e. $\propto 1/\gamma$, leading to a $1/f$ PSD \cite*{Schriefl2006,Paladino2014}.
\begin{align}
S_{1/f}(\omega) \propto \int_0^\infty \frac{d \gamma}{\gamma} \frac{2 (\hbar v)^2 \gamma}{\omega^2 + \gamma^2} = \frac{\pi (\hbar v)^2}{|\omega|}
\label{eq:1f}
\end{align}
The more general case of $P \propto 1/\gamma^\alpha$ leads to a PSD $\propto 1/|\omega|^\alpha$.

However, this PSD leads to a divergent total power when integrated over $\omega$. This is expected because a physical switching rate cannot be infinitely large (which implies charges moving at infinite speeds),
and generally, TLF rates would also be bounded strictly by, e.g., the allowed depth of trap potentials in the material.
A physically more realistic situation is to consider TLFs with rates distributed in a finite range $\gamma \in [\gamma_\text{ir}, \gamma_c]$ \cite*{Shnirman2005}, leading to
\begin{align}
\label{eq:PSD}
S_b(\omega,v,\gamma_\text{ir},\gamma_c) = &\int_{\gamma_\text{ir}}^{\gamma_c} \frac{d \gamma}{\gamma} \frac{2 \gamma (\hbar v)^2}{\omega^2 + \gamma^2} \\
=& 2 (\hbar v)^2 \frac{\arctan(\gamma_c/\omega) - \arctan(\gamma_\text{ir}/\omega)}{\omega}
\nonumber
\end{align}
Qualitatively, this PSD corresponds to a flat region at small frequencies $\omega \lesssim \gamma_\text{ir}$, a $\sim 1/\omega$ decay in the intermediate region $\gamma_\text{ir} \lesssim \omega \lesssim \gamma_c$, and a $\sim 1/\omega^2$ tail for $\omega \gtrsim \gamma_c$, with smooth transitions between regimes.

A common way of probing the noise is to measure the decay of a qubit's coherence, defined as the average of the off-diagonal element of qubit's density matrix
\begin{align}
W(t) = e^{-\chi(t)} = \langle e^{-i \phi(t)} \rangle = \left\langle \frac{\rho_{\uparrow \downarrow}(t)}{\rho_{\uparrow \downarrow}(0)} \right\rangle,
\end{align}
under different pulse sequences. When the statistics of the noise is Gaussian, which is the case in the limit of a large number of TLFs \cite*{Ramon2015}, the coherence function $W(t)$ can be written in terms of the autocorrelation function as
\begin{align}
\chi(t) =  \frac{1}{2} \langle \phi^2(t) \rangle  = \frac{1}{2}  \int_{-\infty}^{\infty} \frac{d\omega}{2\pi \hbar^2} S(\omega) \frac{F(\omega t )}{\omega^2}
\end{align}
where $F(\omega t)$ is called the filter function \cite*{Cywinski2008} and depends on the time-dependent control Hamiltonian or the pulse sequence. The $T_2$ time of a pulse sequence is defined by the relation $\chi(T_2) = 1$  \cite*{Cywinski2008}.

In experiments, noise slower than the timescale on which the experiment is carried out can be calibrated away, which effectively introduces a low frequency cutoff to the correlation function integral as
\begin{align}
\chi(t) =  \int_{\omega_\text{ir}}^{\infty} \frac{d\omega}{2\pi \hbar^2} S(\omega) \frac{F(\omega t )}{\omega^2},
\label{eq:chi}
\end{align}
where $\omega_\text{ir} = 2\pi/T_\text{experiment}$ and we used $S(-\omega) = S^*(\omega) = S(\omega)$ \cite*{Szankowski2017}.

Carr-Purcell-Meiboom-Gill (CPMG)  is a commonly used pulse sequence in noise spectroscopy \cite*{Carr1954,Meiboom1958,Vandersypen2005}. It consists of applying $\sigma_x$ or $\sigma_y$ $\pi$-pulses evenly spaced in time to a system evolving under the Hamiltonian $H_0 = \Omega \sigma_z/2$ otherwise. For an even number of $\pi$-pulses $n$, its filter function is given by \cite*{Cywinski2008}
\begin{align}
F_n^\text{(CPMG)}(\omega t) = 16 \frac{   \sin^4\left(\frac{\omega t}{4n}\right) \sin^2\left(\frac{\omega t}{2}\right)  }{  \cos^2\left(\frac{\omega t}{2 n}\right)  },
\label{eq:Fexact}
\end{align}
where $t$ denotes the total time interval the $\pi$-pulses are applied. This filter function can be approximated using a rectangular wave
\begin{align}
F_n^\text{(CPMG)}(\omega t) \approx 4 n^2 [\Theta(u_n) - \Theta(u_n-2\pi) ]
\label{eq:Fapprox}
\end{align}
where $u_n = (\omega t \mod 2\pi n) - \pi (n-1)$ and $\Theta(x)$ is the Heaviside step function.
Using this approximate filter function with $S(\omega)/\hbar^2 = A_0^{1+\alpha} /\omega^\alpha$ PSD, the exponent of the coherence function can be calculated as
\begin{align}
\chi(t) \approx \frac{C_\alpha}{2\pi}\frac{(A_0 t)^{1+\alpha}}{n^\alpha}
\end{align}
where $C_\alpha = \pi^{-2} (2\pi)^{1-\alpha} \sum_{k=1}^\infty (k-1/2)^{-(\alpha+2)}$ or $C_\alpha = 4 (2^{\alpha+2}-1) \zeta(\alpha+2)/(2\pi)^{1+\alpha}$ with $\zeta(x)$ as the Riemann zeta function, and this yields $C_1 \approx 0.85$ for $\alpha = 1$ \cite*{Cywinski2008}.  This leads to the  decay rate and scaling relation \cite*{Cywinski2008,Medford2012}
\begin{align}
\chi(t) = \left(\frac{t}{T_2}\right)^{1+\alpha}, \qquad T_2 = T_2^0 n^{\frac{\alpha}{1+\alpha}}
\label{eq:chi-1f}
\end{align}
where $1/T_2^0 = \left(C_\alpha/2\pi\right) ^\frac{1}{1+\alpha}A_0$.
We remark that the corresponding PSD here assumes that $\alpha$ and $A_0$ are constants at all frequencies $\omega$ which contribute to $\chi(t)$ in Eq~(\ref{eq:chi}).

For CPMG pulses with high enough number of $\pi$-pulses $n$ (such that $t/n  \ll \gamma^{-1},v^{-1}$ for short times \cite*{Szankowski2017}, as well as in the weak coupling limit of $v \ll \gamma$ in general \cite*{Cywinski2008}), non-Gaussian effects are suppressed for timescales of interest $\sim T_2$. Non-gaussian effects are also suppressed in limit of a large number of TLFs \cite*{Ramon2015} which we assume here, although $n$ dependence of such corrections is weak in the weak coupling limit \cite*{Szankowski2017}. A more careful analysis of the limits of the Gaussian approximation can be found in Ref.~\onlinecite*{Szankowski2017}.

\emph{Noise spectroscopy experiment in Si/SiGe.}
A silicon spin qubit in a Si/SiGe or SiMOS quantum dot can be described by the following effective spin Hamiltonian
\begin{align}
H = \frac{1}{2} g \mu_B \boldsymbol B \cdot \boldsymbol \sigma
\end{align}
where $\boldsymbol B$ is the magnetic field felt by the electron at its position within the dot \cite*{Tokura2006}. In general, the magnetic field is not spatially uniform, so its value depends on the electron's position in the dot. Since the electron's position can be controlled by the electric field, the magnetic field gradient allows electrical control of the spin qubit. This, however, also makes the qubit susceptible to fluctuations in the electric field caused by charge noise, which can be caused, e.g., by other electrons hopping between nearby defects or dangling bonds \cite*{Culcer2009,Ramon2010}.

Another important source of noise spin qubits in silicon is the random flips of the remnant $^{29}$Si nuclei with nuclear spin. The experiment \cite*{Yoneda2017a} uses isotopically purified $^{28}$Si with no nuclear spin with a remnant $^{29}$Si concentration of 800ppm, so hyperfine noise can still impact the PSD. However, the PSD of this noise decays much faster than the charge noise, thus charge noise is expected to dominate the high frequency noise.

\begin{figure}
\includegraphics[width=0.49\columnwidth]{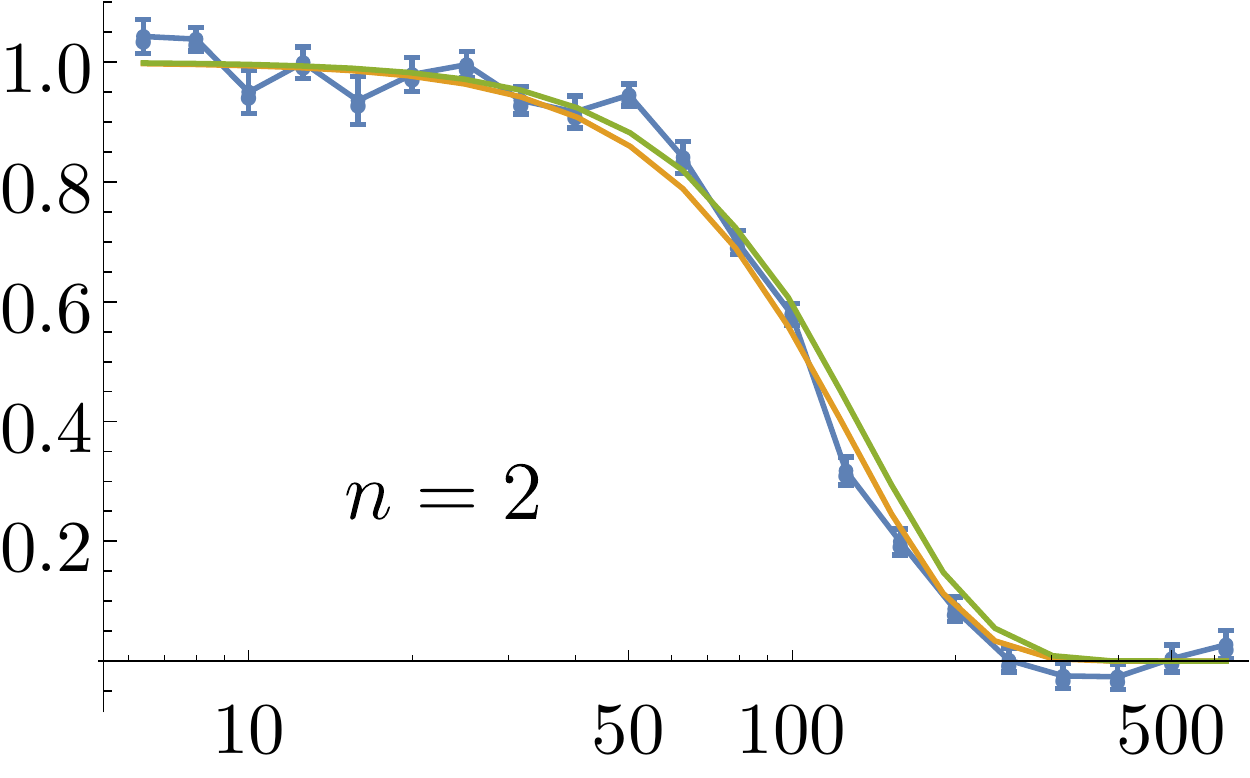}
\includegraphics[width=0.49\columnwidth]{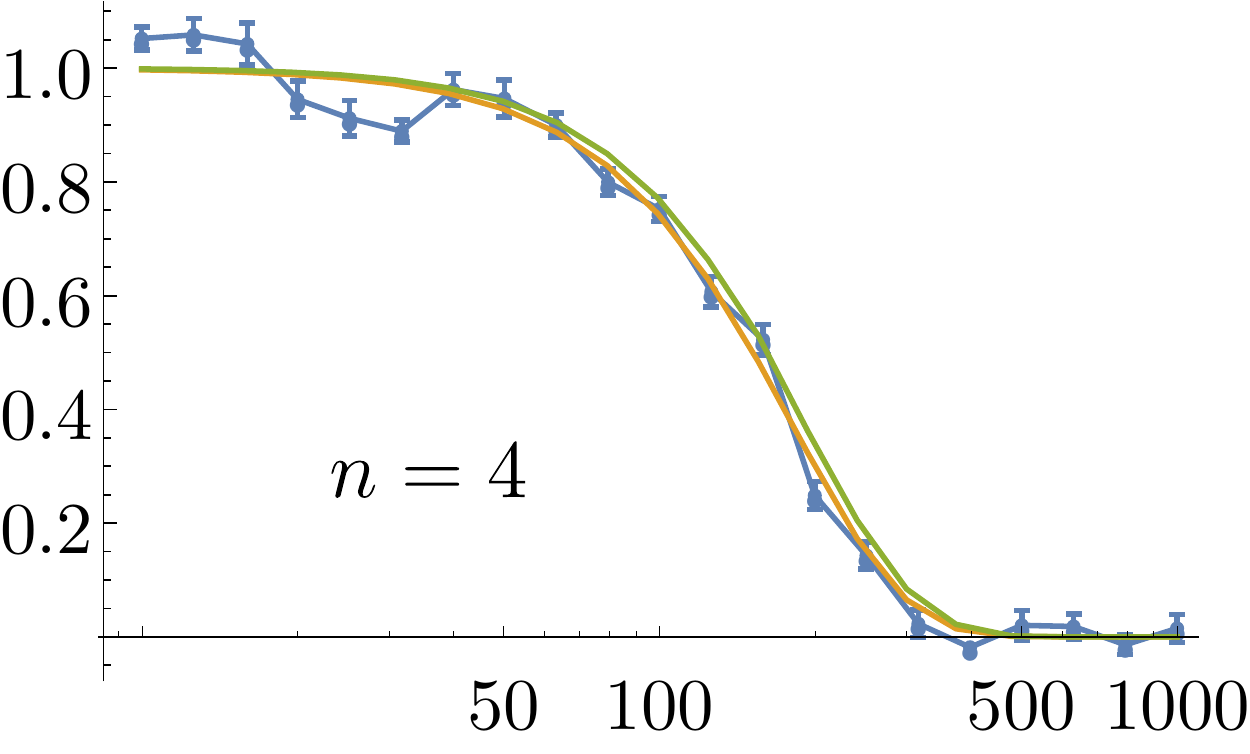}
\includegraphics[width=0.49\columnwidth]{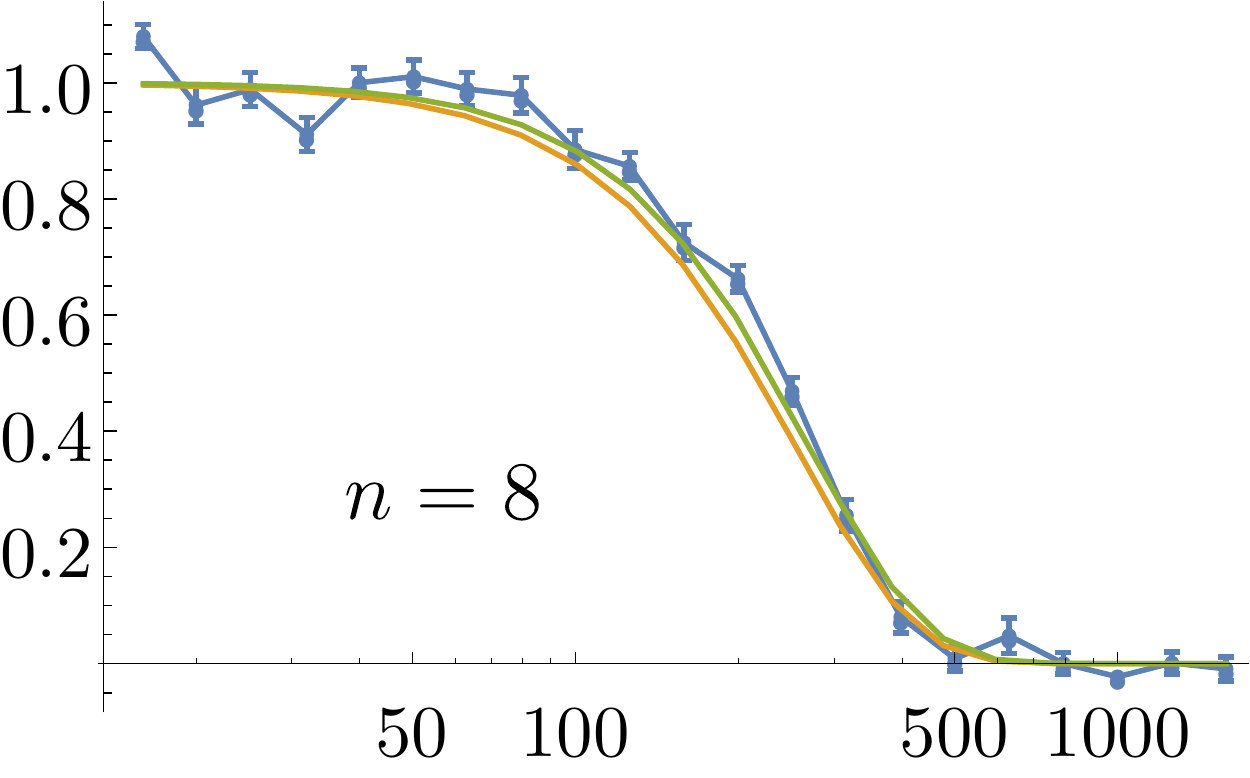}
\includegraphics[width=0.49\columnwidth]{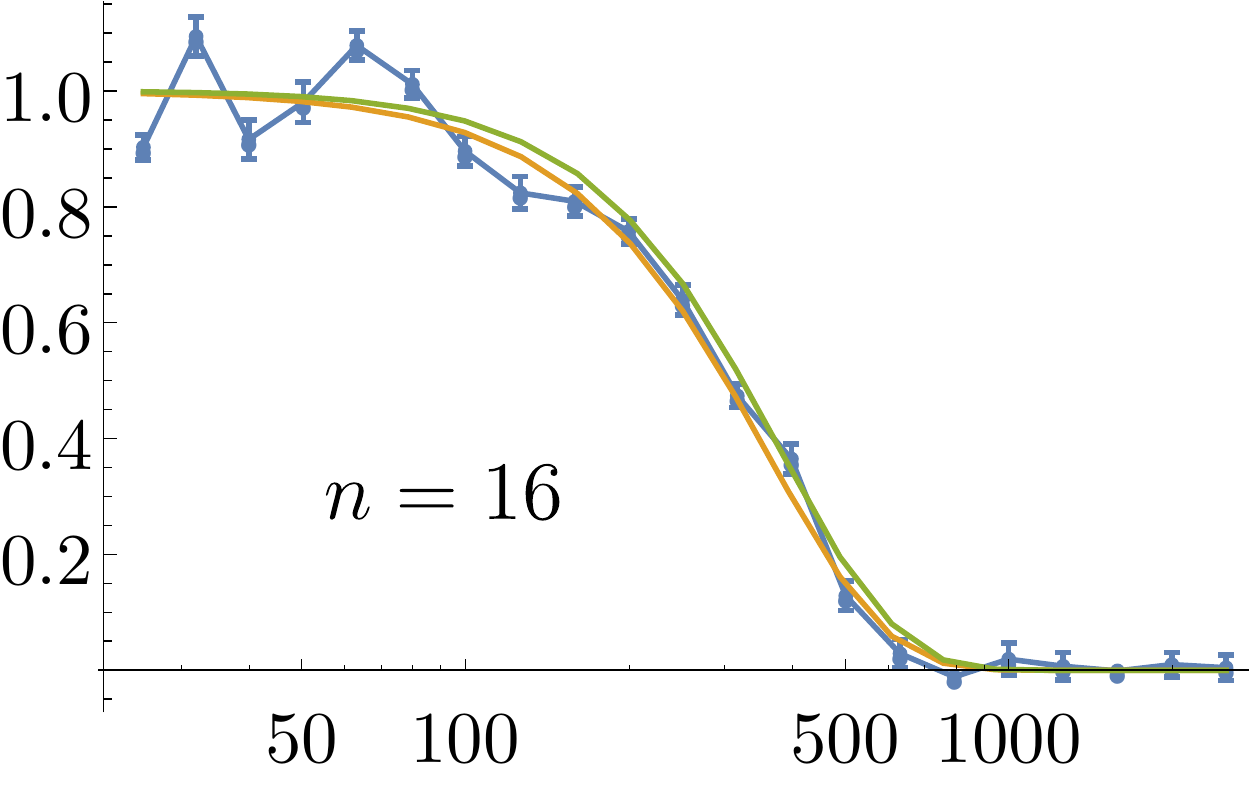}
\includegraphics[width=0.49\columnwidth]{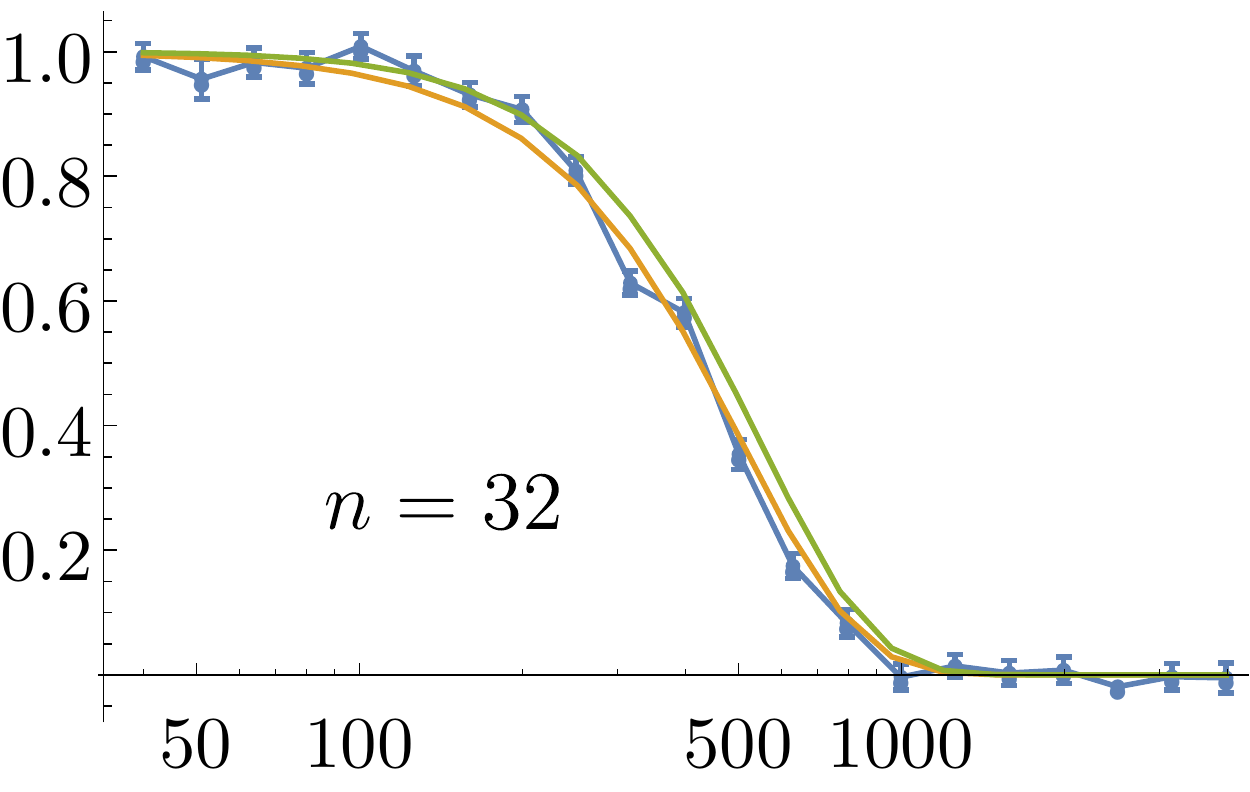}
\includegraphics[width=0.49\columnwidth]{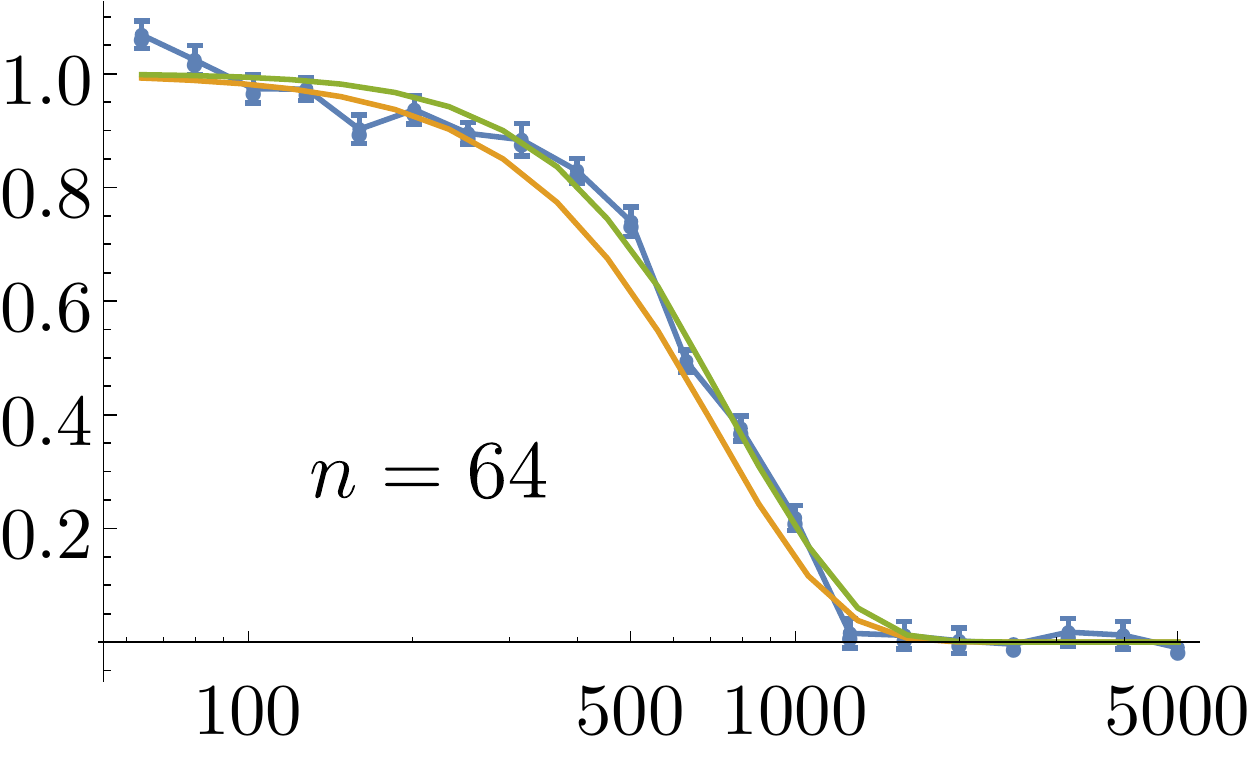}
\includegraphics[width=0.49\columnwidth]{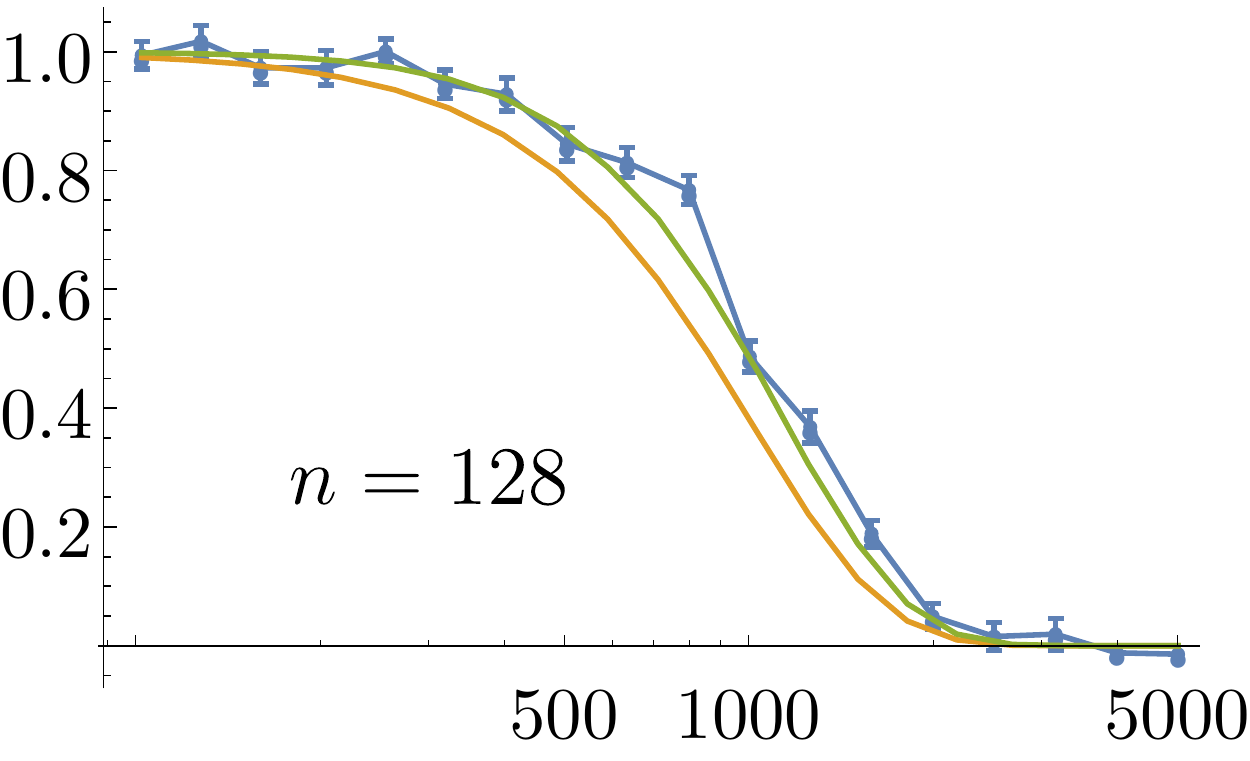}
\includegraphics[width=0.49\columnwidth]{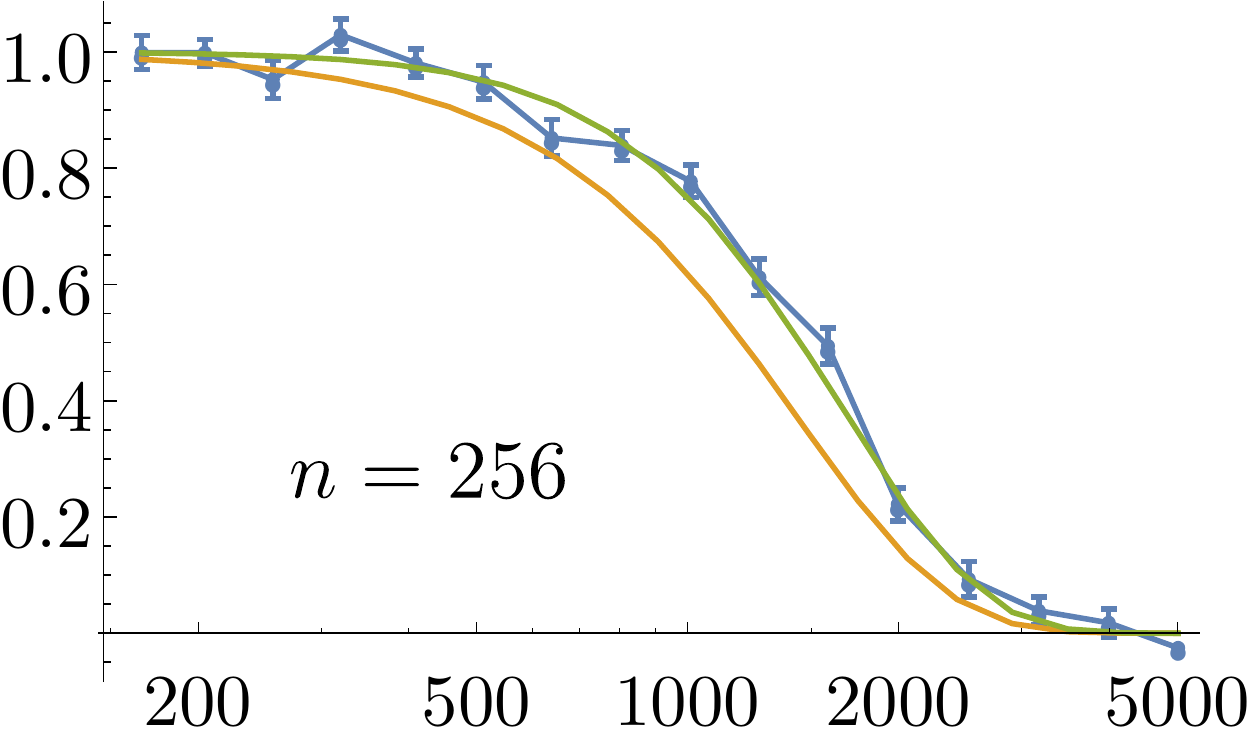}
\includegraphics[width=0.49\columnwidth]{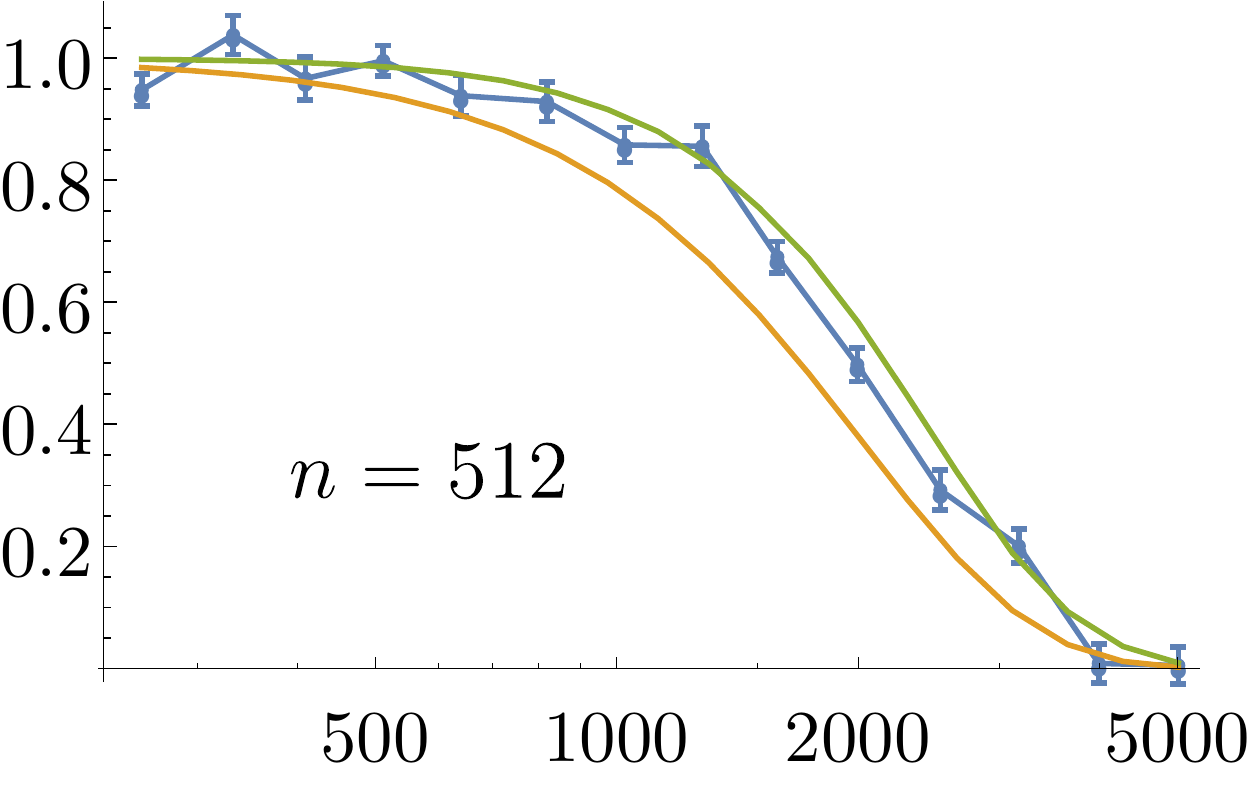}
\includegraphics[width=0.49\columnwidth]{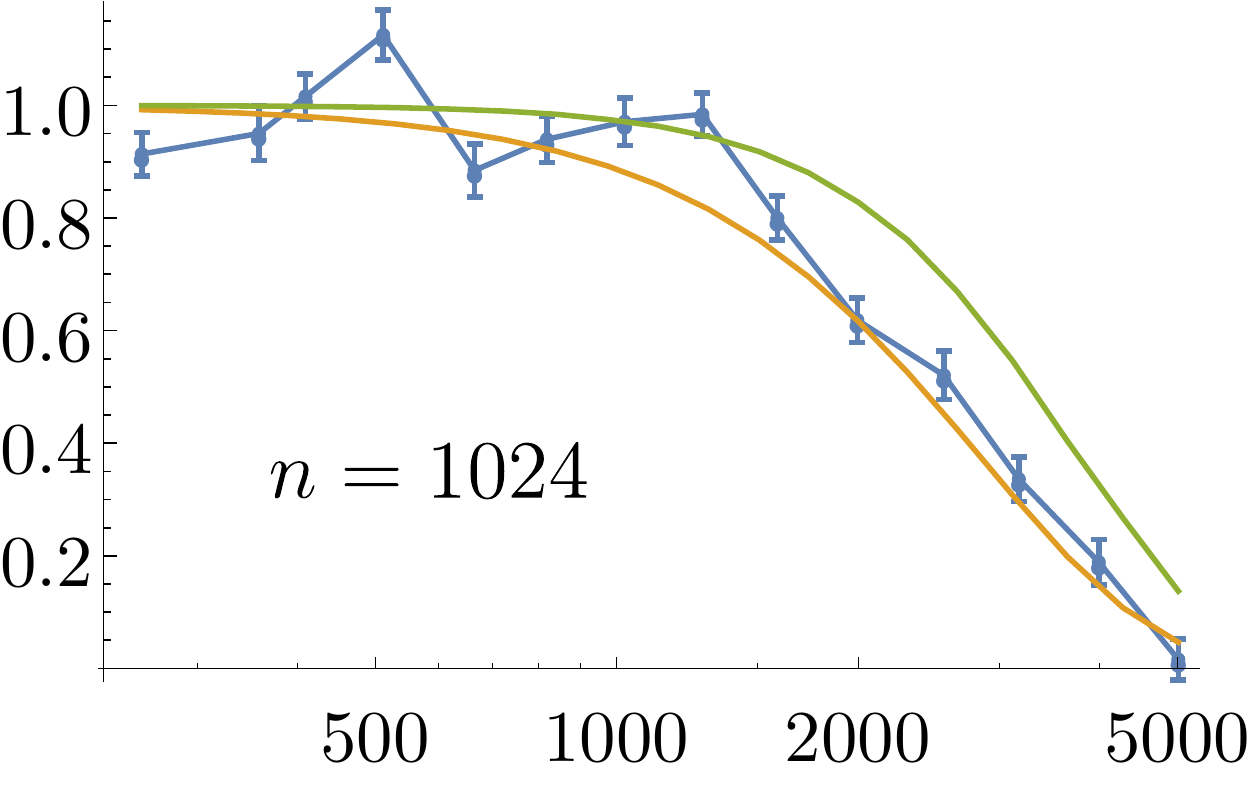}
\caption{(Color online) Coherence function $W(t)$ as a function of total CPMG time $t$ in $\mu$s, for CPMG different number $\pi$-pulses $n$. The blue line segments correspond to joined experimental $W(t)$ data points, orange dashed curves correspond to $W(t) = \exp(-[t/T_2(n)]^\alpha)$ predicted by the PSD $\hbar^2 A_0^{1+\alpha}/\omega^\alpha$ with $\alpha = 1.01$ and $T_2^0 \approx 91 \mu$s, green curves correspond to $W(t)$ corresponding to the PSD given in Eq.~\ref{eq:PSD} with $v/2\pi \approx 2.65$kHz, $\gamma_\text{ir}/2\pi = 0.1$kHz and $\gamma_c/2\pi = 200$kHz.}
\label{fig:cpmg}
\end{figure}

The blue line segments in Fig.~\ref{fig:cpmg} show the joined experimental data points from  CPMG experiments \cite*{Yoneda2017a}, with 1$\sigma$ error bars obtained from experimental data \cite{Yoneda2019}. Each subfigure corresponds to a CPMG experiment with a different number of $\pi$ pulses, $n$. For a given $n$, the spin is initialized in the spin down state followed by a $\pi/2$ rotation around $\sigma_x$. After applying an evenly spaced sequence of $\pi$-pulses around $\sigma_y$ in a period of time $t$, another $\pi/2$ pulse is applied around and axis that is swept in the $xy$-plane (in order to counter systematic $\sigma_z$ errors) before the readout. Each data point shows the average spin up probability after the readout corresponding to $\chi(t)$.

\emph{Noise models.}
\label{sec:results}
To understand the underlying PSD of the device from the experiment \cite*{Yoneda2017a}, we compare the experimental coherence function data obtained from CPMG experiment, against the curves predicted by $S_{1/f}(\omega)$ from Eq.~\ref{eq:1f} with fixed $\alpha$ and $A_0$, and $S_b(\omega,v,\gamma_\text{ir},\gamma_c)$ from Eq.~\ref{eq:PSD}.

The dashed orange curves in Fig.~\ref{fig:cpmg} shows the coherence function corresponding to $S(\omega)/\hbar^2 = A_0^{1+\alpha} /\omega^\alpha$, using Eq.~\ref{eq:chi-1f}, with exponent $\alpha = 1.01$ from Ref.~\onlinecite*{Yoneda2017a} and the decay time $T_2^0 \approx 91\mu$s obtained by fitting to the CPMG data \footnote{We used local optimization to fit to the noisy experimental data with an initial guess which approximates to the $1/f^{1.01}$ spectrum provided in Ref. \onlinecite*{Yoneda2017a} within the frequency range probed by the CPMG experiments, and imposed constraints on cutoff frequencies to avoid frequencies not reliably probed by the CPMG filter function.} \footnote{We numerically checked that using the CPMG filter function Eq.~\ref{eq:Fexact} rather than the approximate Eq.~\ref{eq:Fapprox} effectively renormalizes $T_2^0$ as $\approx 85\mu$s after a fit to data, and results in visually indistinguishable decay curves.}. This simple noise model is a remarkably good fit to the experimental data for all $n$, as shown in Fig.~\ref{fig:cpmg}, and especially for $n=2$ up to $n=32$. However, we note that from $n=64$ to $n=512$, this simple model starts to drift away from the experimental data in a consistent manner by predicting a faster decay with increasing $n$. (The $n=1024$ data breaks this trend, and has other puzzling features as well, but we will discuss this later.)  This hints at the possibility of extracting from the data not only the $1/f$-like behavior of the PSD beautifully demonstrated in Ref.~\onlinecite{Yoneda2017a} across 7 decades of frequency, but also the cutoff frequency at which it starts transitioning to $1/f^2$-like behavior.

\begin{figure}
	\includegraphics[width=0.9\columnwidth]{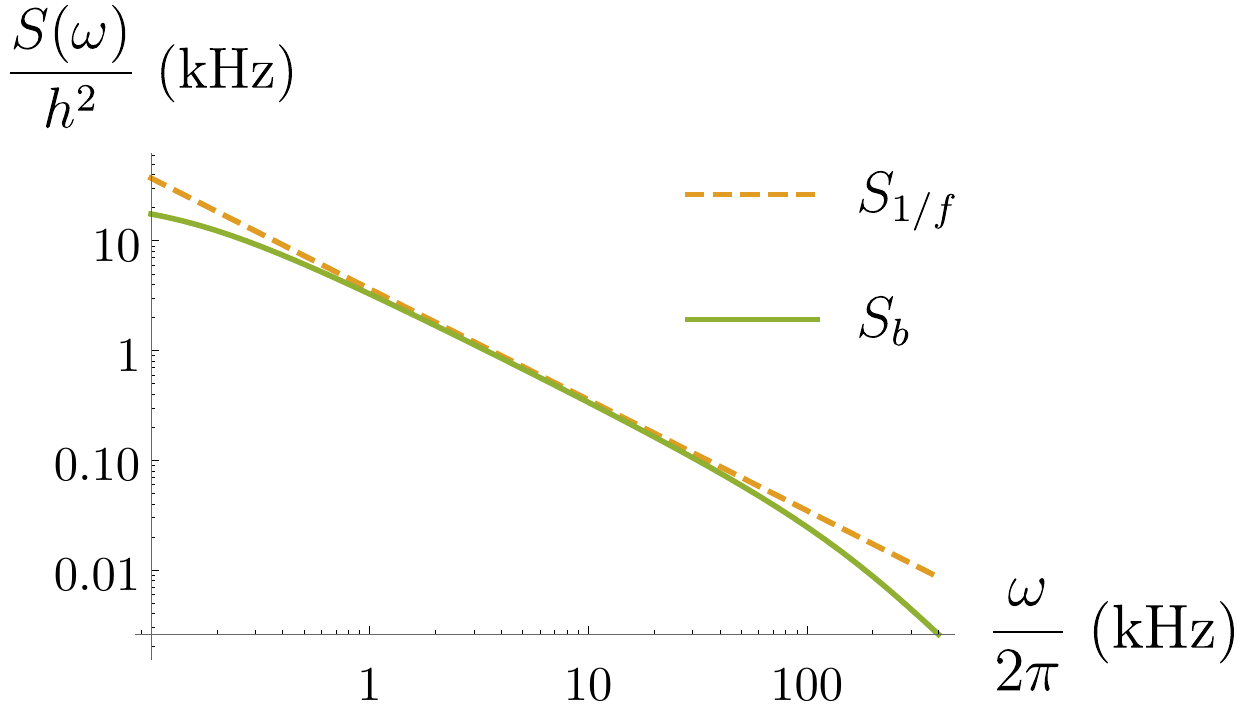}
\caption{(Color online) Comparison of fitted PSD functions used in Fig.~\ref{fig:cpmg} in a frequency range from 0.1kHz to 400kHz. At the TLF cutoff frequency $\omega/2\pi = \gamma_c/2\pi=200$kHz, $S_b$ becomes smaller than $S_{1/f}$ by a factor of $\approx 2$.}
\label{fig:psd}
\end{figure}

Clearly, the relations given in Eq.~\ref{eq:chi-1f} are derived under the assumption that the noise spectrum is $\hbar^2 A_0^{1+\alpha}/\omega^\alpha$ up to frequencies much higher than the time scale defined by characteristic decay times, and the power spectrum inferred using it from the experimental data can be self-consistent if and only if the decoherence decays agree with Eq.~\ref{eq:chi-1f}, and the inferred power spectrum obeys a power-law with constant exponent and amplitude in a wide enough region. When fitting to Eq.~\ref{eq:chi-1f} requires $\alpha$ to change as a function of $\omega$, the theory would no longer be self-consistent since $\chi(t)$ was derived under the assumption that $\alpha$ and $A_0$ do not depend on $\omega$ in the region of interest. Another issue with $\omega$-dependent $\alpha$ or $A_0$ is that the filter function of a CPMG sequence is not perfectly localized around a certain frequency, but is rather a frequency comb probing higher harmonics as well (Slepians, which are well localized in both time and frequency domains, can be used to eliminate this issue \cite*{Frey2017}). Therefore, explaining a deviation from the $A_0^{1+\alpha}/f^\alpha$ model with constant $\alpha$ and $A_0$ at a certain frequency $\omega_0$ in terms of a local effect $\{\alpha,A_0\} \to \{\alpha(\omega),A_0(\omega)\}|_{\omega=\omega_0}$ may not be accurate.

The consistent character of the exponent deviations in Ref.~\onlinecite{Yoneda2017a} suggests that the decay of the underlying PSD gradually becomes faster than $1/f^{1.01}$ with increasing frequency. Regardless of the choice for $\alpha$ and $A_0$, this drifting feature of $\alpha$ or $A_0$ remains consistent as the frequency is varied, rendering any constant choice for these parameters an inadequate fit.

We note that non-Gaussian corrections cannot explain this mismatch, since the observed mismatch gets stronger with increasing $n$. The timescale of CPMG experiments increase as $\propto n^{\frac{\alpha}{\alpha+1}}$ while non-Gaussian effects are suppressed or remain small \cite*{Ramon2015}, which means such a mismatch would decrease with increasing $n$.

On the other hand, a gradual increase in the exponent from $\alpha=1$ is consistent with a PSD obtained from an ensemble of TLFs over a finite range of characteristic switching frequencies, as given in Eq.~\ref{eq:PSD}. Such a PSD has three independent parameters: $\gamma_\text{ir}$ ($\gamma_c$) impacts the fitting only at lower (higher) $n$ and $v$ determines its overall strength. This locality is due to
the fact that $T_2$ time grows with $n$, and the most significant contribution of the CPMG filter function comes from its first peak around $\omega t = n \pi$ as the contributions from later peaks are suppressed by the factor $1/\omega^2$ in Eq.~\ref{eq:chi}. Using the physical model of $S(\omega)$ from Eq.~\eqref{eq:PSD} with the fitting parameters $\gamma_\text{ir}/2\pi= 0.1$kHz, $\gamma_c/2\pi = 200$kHz, $2\pi/v \approx 379\mu$s,
we integrate Eq.~\ref{eq:chi} from $\omega_\text{ir}/2\pi = 1/T_\text{experiment} = 1/100$s, which yields a good fit for experimental data for CPMG data with $n$ from $2$ to $512$ as shown by the green curves in Fig.~\ref{fig:cpmg}. The value of $\gamma_\text{ir}$ is not important in this context, and could be chosen arbitrarily smaller without affecting the fit, since the CPMG filter functions are not sensitive to the PSD at very low frequency.  However, as discussed above, the value of $\gamma_c$ is very meaningful and important. The error bars allow $\gamma_c/2\pi$ to be between $\approx 170$kHz and $\approx 240$kHz.  A comparison of this PSD to the $1/f^{1.01}$ fit is shown in Fig.~\ref{fig:psd}. We remark here that these fitting parameters also satisfy the Gaussian approximation, since fast fluctuators fall into the weak coupling regime as $v \approx 2$kHz is much smaller than fast TLFs $\gamma \sim 100$kHz, and slow fluctuators $\gamma \lesssim 2$kHz satisfy $t/n \ll v^{-1}$.

However, this model does not match the $n=1024$ data.
This might be due to the presence of a TLF at a frequency around $\sim 1$MHz. However, we note that the $n=1024$ data appears to be more noisy, which might have lead to an error in estimating the readout visibility range and a rescaling of the data might be required. On the other hand, the slope of the $1/f^{1.01}$ curve at the tail does not match the experimental data either. Additional CPMG measurements up to $n=4096$, or Slepian-shaped pulses \cite*{Frey2017} with envelope modulation at around $\sim 1$MHz could be used to definitively determine the PSD at frequencies $\gtrsim 200$kHz.

\emph{Conclusion.}
We have analyzed the coherence decay data of a spin qubit from a recent experiment in Si/SiGe \cite*{Yoneda2017a}, and found that deviations from a $1/f^\alpha$ noise spectrum model follow a consistent trend indicating a region where $\alpha$ has to grow with frequency. We have shown that an ensemble of TLFs with a log-linear distribution of switching frequencies in a finite range can provide a better fit to the experimental data sets \cite*{Yoneda2017a}, which are systematically consistent with a soft cutoff frequency at 200kHz, near the upper end of the probed frequencies.  Unfortunately, the one exception to this is the longest CPMG data set, which probes around 300kHz and which remains consistent with $1/f$ noise, and one can speculate that might be due to an additional TLF with switching rate around $\sim 1$MHz, but a conclusive answer requires probing even higher frequencies.  Measurements of the noise spectrum in silicon devices hosting spin qubits have so far been limited to sub-MHz frequencies \cite*{Kawakami2016,Yoneda2017a,Chan2018a}, and measurements probing up to 1MHz and above are necessary to pin down exactly how the PSD behaves at frequencies relevant for robust gate design \cite*{Gungordu2018a}, but our results provide hope that the all-important charge noise rolloff may be occurring at frequencies just beyond the currently measured range.

\emph{Acknowledgements.}
We thank Jun Yoneda for sharing the experimental data with us, and useful discussions. This research was sponsored by the Army Research Office (ARO), and was accomplished under Grant Number W911NF-17-1-0287.

\bibliography{siqd,extra}

\end{document}